\documentclass[11pt]{article}
\usepackage{fullpage}
\usepackage{amsmath}
\usepackage{amsfonts}
\usepackage[dvips]{epsfig}

\def\##1{\underline{#1}}
\def\=#1{\underline{\underline{#1}}}

\def\+#1{\underline{\bf #1}}
\def\*#1{\underline{\underline{\bf #1}}}

\def\.{\mbox{ \tiny{$^\bullet$} }}

\def\eps{\epsilon}
\def\epso{\epsilon_{\scriptscriptstyle 0}}
\def\muo{\mu_{\scriptscriptstyle 0}}
\def\etao{\eta_{\scriptscriptstyle 0}}
\def\ko{k_{\scriptscriptstyle 0}}
\def\omegao{\omega_{\scriptscriptstyle 0}}

\def\c#1{\cite{#1}}
\def\l#1{\label{#1}}
\def\r#1{(\ref{#1})}

\def\le{\left(}
\def\ri{\right)}
\def\les{\left[}
\def\ris{\right]}
\def\lec{\left\{}
\def\ric{\right\}}

\renewcommand{\thefootnote}{\fnsymbol{footnote}}

\begin{document}

\begin{center}

Revised on October 3, 2003\\

\LARGE{ {\bf
Plane waves with
negative phase velocity\\  in Faraday chiral mediums
}}
\end{center}

\begin{center}

\bigskip
Tom G. Mackay\footnote{Fax: +44 131 650
6553; e--mail: T.Mackay@ed.ac.uk}\\

{\em School of Mathematics, University of Edinburgh \\
     James Clerk Maxwell Building, The King's Buildings\\
      Edinburgh  EH9 3JZ, United Kingdom}
\bigskip

Akhlesh Lakhtakia\footnote{Fax: +1 814 863 7967;
e--mail: axl4@psu.edu}\\

{\em CATMAS --- Computational and Theoretical Materials Sciences Group \\
     Department of Engineering Science and Mechanics \\
     Pennsylvania State University, University Park, PA
     16802--6812, USA}

\end{center}

\renewcommand{\thefootnote}{\arabic{footnote}}
\setcounter{footnote}{0}

\bigskip

\noindent {\bf Abstract.}
The propagation of plane waves in a  Faraday chiral medium is
investigated.
Conditions for the phase velocity
to be directed opposite to the direction of power flow are derived
for propagation in an arbitrary direction;
simplified conditions which apply to propagation
parallel to the distinguished axis are also established.
These negative phase--velocity conditions are explored numerically using a
representative  Faraday chiral medium, arising from the
  homogenization of an isotropic chiral medium and a
magnetically biased
ferrite. It is demonstrated that the  phase velocity
may be directed opposite to power flow, provided that the
gyrotropic
  parameter of the ferrite component medium is sufficiently large
compared
with the corresponding
nongyrotropic permeability parameters.

\bigskip
\noindent
PACS number(s): 41.20.Jb, 42.25.Bs, 83.80.Ab

\section{Introduction}

Homogeneous
mediums which support the propagation of waves with phase velocity
directed opposite to the direction of power flow
have attracted much attention lately \c{Ziolkowski, LMW02, Veselago03}.
The archetypal example of such a medium is the isotropic
dielectric--magnetic medium
with simultaneously negative real permittivity and
negative real permeability scalars, as first described
Veselago in  the late 1960s \c{Veselago68}. A range of exotic and
potentially useful electromagnetic phenomenons, such as negative refraction,
inverse Doppler  shift, and inverse \u{C}erenkov radiation, were
predicted for this type of medium
\c{Veselago03, Veselago68}.
Recent experimental studies involving the microwave illumination
of certain composite
metamaterials  \c{Smith00, Shelby01}~---~which
followed on from
earlier works of
Pendry \emph{et al.} \c{Pendry98, Pendry99}~---~are supportive of
Veselago's
predictions and have prompted an intensification
of  interest in this area. In particular, a general
condition~---~applicable to dissipative isotropic
dielectric--magnetic mediums~---~has been
derived for  the phase velocity to be  directed opposite to power flow
\c{MartinLW02}; and this condition shows that the real parts of both
the permittivity and the
permeability scalars do not have to be negative.

A consensus has yet to be reached on terminology. For the present
purposes, a medium supporting wave propagation with phase velocity
directed opposite to  power flow is  most aptly
referred to as a \emph{negative phase--velocity} medium. However, the
reader
is alerted  that alternative terms, such as left--handed
material \c{Veselago03}, backward medium \c{Lindell01},
double--negative medium \c{Ziolkowski},
and negative--index medium \c{Valanju}, are also in circulation. A
discussion of this
  issue is available elsewhere \c{LMW03}.

The scope for the
phase velocity to be  directed opposite to power flow may be greatly
extended by considering non--isotropic mediums, as has been indicated
by  considerations of
uniaxial dielectric--magnetic mediums \c{Lindell01, Hu, Kark}. The focus of the present
  communication is  the propagation of negative phase--velocity plane waves
  in  \emph{Faraday chiral mediums} (FCMs)
  \c{Engheta92, WL98}. These mediums combine natural
optical activity~---~as exhibited by isotropic chiral mediums
\c{Beltrami}~---~with
Faraday rotation~---~as exhibited by gyrotropic mediums
\c{Lax,Chen,Coll}. A FCM  may be
  theoretically conceptualized as a
homogenized composite medium (HCM) arising  from the blending together of
an isotropic
chiral  medium  with either a magnetically biased ferrite
\c{WLM98} or a
magnetically biased
plasma \c{WM00}. The HCM  component mediums are
envisioned as random particulate distributions.
The homogenization process is justified provided that the particulate
length scales in the mixture of components are small compared with 
electromagnetic wavelengths. A vast literature on the estimation
of constitutive parameters of HCMs exists;
see Refs. \c{L96, M03}, for example.
The constitutive relations of FCMs have
been rigorously established for some time \c{WL98}, although inappropriate
use still occurs  \c{Fu}.

In the following sections, wavenumbers and corresponding
electric field phasors are delineated from eigenvalue/vector analysis  for
planewave
propagation in an arbitrary
 direction.
Simplified expressions for
these quantities are derived for propagation parallel to
the biasing (quasi)--magnetostatic field \cite[p. 71]{Chen}. A
general condition for the phase velocity to be directed opposite to
power flow
  is established. The theoretical
analysis is illustrated by means of a representative
numerical example: the constitutive parameters
of  FCMs arising from a specific homogenization scenario
are estimated and then used
  to explore the wave propagation characteristics.

As regards notation, vectors are  underlined whereas dyadics are
double underlined.
All electromagnetic field phasors and constitutive parameters  depend
implicitly on the circular frequency $\omega$ of the electromagnetic field.
Unit vectors are
denoted by the  superscript $\, \hat{} \,$ symbol,
  while $\=I = \hat{\#x}\,\hat{\#x} +
  \hat{\#y}\,\hat{\#y} +  \hat{\#z}\,\hat{\#z}$ is the identity dyadic.
The complex conjugate of a quantity $q$ is written as $q^*$; the real
part of $q$ is written as $\mbox{Re} \, \{ q \}$.
The free--space (i.e., vacuum)
wavenumber is denoted by $\ko = \omega \sqrt{\epso \muo}$ where
$\epso$ and $\muo$ are the permittivity and permeability of free
space, respectively; and $\etao = \sqrt{\muo / \epso}$ represents the
intrinsic impedance of free space.

\section{Analysis}

\subsection{Preliminaries}

The propagation of plane waves with field phasors
\begin{equation}
\left.\begin{array}{l}
\#E(\#r) = \#E_0\, \exp (i \ko \tilde{k} \, \hat{\#u} \. \#r )\\[5pt]
\#H(\#r) = \#H_0\, \exp (i \ko \tilde{k} \, \hat{\#u} \. \#r )
\end{array}\right\}
\l{pw}
\end{equation}
in a
FCM is considered.
Such a medium is  characterized
  by  the frequency--domain constitutive relations \c{WL98}
\begin{equation}
\left.
\begin{array}{l}
  \#D (\#r) = \=\eps\.\#E (\#r) + \=\xi\.\#H (\#r) \,\\[5pt]
  \#B (\#r) = - \=\xi\.\#E (\#r) + \=\mu\.\#H (\#r) \, \l{FCM_cr}
\end{array}
\right\},
\end{equation}
with constitutive dyadics
\begin{equation}
\left.
\begin{array}{l}
  \=\eps =
\epso \les \, \eps \, \=I
- i \eps_g \, \hat{\#z} \times \=I +
\le \, \eps_z - \eps \, \ri \,  \hat{\#z}\,  \hat{\#z} \, \ris\\
\vspace{-2mm} \\
  \=\xi =
i \sqrt{\epso \muo} \, \les \, \xi \, \=I
- i \xi_g \, \hat{\#z} \times \=I +
\le \, \xi_z - \xi \, \ri \,
  \hat{\#z} \,  \hat{\#z}
\ris\\
\vspace{-2mm} \\
  \=\mu =
\muo \les \, \mu \, \=I
- i \mu_g \, \hat{\#z} \times \=I +
\le \, \mu_z - \mu \, \ri \,  \hat{\#z}\,  \hat{\#z} \, \ris
\end{array}
\right\}. \l{FCM}
\end{equation}
Thus, the distinguished axis of the FCM is chosen to be the $z$ axis.
For FCMs which arise as  HCMs, the gyrotropic
parameters $\eps_g$, $\xi_g$ and $\mu_g$ in \r{FCM} develop due to the
gyrotropic properties of the ferrite or plasma component mediums.
Parenthetically, it is  remarked
  that more general FCMs can develop
  through the homogenization of component mediums based
on nonspherical particulate geometries \c{WM00, MLW01a}.

In general, the
relative wavenumber $  \tilde{k} $ in \r{pw}
is complex valued;
 i.e.,
\begin{equation}
\tilde{k} = \tilde{k}_R + i \tilde{k}_I, \qquad (\tilde{k}_R,
\tilde{k}_I \in \mathbb{R}).
\end{equation}
It
may be  calculated from  the
   planewave dispersion relation
\begin{eqnarray}
\mbox{det} \, \les \, \=L(i \ko  \tilde{k} \, \hat{\#u}) \, \ris  = 0
\,, \l{disp}
\end{eqnarray}
which arises from
  the  vector Helmholtz equation
\begin{eqnarray}
\=L(\nabla) \. \#E (\#r) = \#0\,, \l{Helm}
\end{eqnarray}
wherein
\begin{eqnarray}
\=L(\nabla) = \le \nabla \times \=I + i \omega \=\xi \ri
\. \=\mu^{-1} \. \le \nabla \times \=I + i \omega \=\xi \ri
- \omega^2 \=\eps \,. \l{L_nabla}
\end{eqnarray}

Of particular interest  is the
orientation of the phase velocity, as specified by
the direction of $  \tilde{k}_R \, \hat{\#u}$, relative to the direction of
power  flow given by the  time--averaged
  Poynting vector
$
\#P (\#r) =
\frac{1}{2} \mbox{Re} \, \les \, \#E (\#r) \times \#H^* (\#r) \,\ris
$.
The combination of  the constitutive relations \r{FCM_cr} with
  the source--free Maxwell curl postulates
\begin{equation}
\left.
\begin{array}{l}
\nabla \times \#E (\#r) = i \omega \#B (\#r) \\[5pt]
\nabla \times \#H (\#r) = -i \omega \#D (\#r)
\end{array}
\right\}
\end{equation}
yields
\begin{equation}
 \#P (\#r)  =
 \frac{1}{2} \,  \exp \le - 2 \ko \tilde{k}_I \,  \hat{\#u}\.\#r \ri \,
 \mbox{Re} \, \lec  \#E_0 \times
\les
(\=\mu^{-1})^* \.\le \sqrt{\epso \muo} \tilde{k}^* \,  \hat{\#u} \times \#E^*_0 +
\=\xi^* \. \#E^*_0 \ri \ris \ric  \l{P_pw}
\end{equation}
for plane waves \r{pw}.

In the remainder of this section,
the quantity $  \tilde{k}_R \, \hat{\#u} \. \#P(\#r) $ is
derived for
planewave propagation in an arbitrary direction;
without loss of generality, $\hat{\#u}$ is taken to lie
in the $xz$ plane  (i.e., $\hat{\#u} = \hat{\#x}\, \sin \theta  +  \hat{\#z}\,\cos
\theta  $). Further manipulations reveal the
simple form $  \tilde{k}_R \, \hat{\#u} \. \#P (\#r) $ adopts
for propagation
along the FCM distinguished axis (i.e., $\hat{\#u} = \hat{\#z}$).

\subsection{ Propagation in the $xz$ plane}\label{2.xz}

 For $  \hat{\#u} = \hat{\#x}\, \sin \theta  +  \hat{\#z}\,\cos
\theta  $,  the dispersion relation  \r{disp} may be
represented by
 the
quartic polynomial
\begin{equation}
a_4 \tilde{k}^4 + a_3 \tilde{k}^3 + a_2 \tilde{k}^2 + a_1 \tilde{k} + a_0 =0\,,
\l{quartic}
\end{equation}
with coefficients
\begin{eqnarray}
a_4 &=&
\le \eps \sin^2 \theta + \eps_z \cos^2 \theta \ri \le \mu \sin^2
\theta + \mu_z \cos^2 \theta \ri
- \le  \xi \sin^2 \theta +  \xi_z \cos^2 \theta \ri^2, \\
a_3 &=&
2 \cos \theta \big\{
\sin^2 \theta \les
\mu_g \le \eps \xi_z  - \eps_z \xi \ri + \eps_g \le \mu \xi_z - \mu_z
\xi \ri + \xi_g \le \mu \eps_z + \eps \mu_z - 2 \xi \xi_g \ri \ris
 \nonumber
\\
&&
+ 2 \cos^2 \theta \xi_g \le \eps_z \mu_z - \xi^2_z \ri \big\}, \\
a_2 &=&
\sin^2 \theta \Big\{
\mu \mu_z \le \eps^2_g - \eps^2 \ri + \le \xi^2 + \xi^2_g \ri
\le \mu \eps_z + \eps \mu_z \ri -
2 \xi \les \xi_z \le \xi^2_g - \xi^2 \ri + \mu_g \eps_z \xi_g \ris
\nonumber \\ &&
- 2 \eps_g \les \xi_z \le \mu_g \xi  - \mu \xi_g \ri + \mu_z \xi \xi_g
\ris
- \eps \les \eps_z \le \mu^2 - \mu^2_g \ri + 2 \xi_z \le \mu \xi - \mu_g
\xi_g \ri \ris \Big\} \nonumber \\ &&
+ 2 \cos^2 \theta \le \eps_z \mu_z - \xi^2_z \ri
\le 3 \xi^2_g - \xi^2 - \eps_g \mu_g - \eps \mu \ri , \\
a_1 &=& 4 \cos \theta \le \eps_z \mu_z  - \xi^2_z \ri \Big[ \xi
\le \eps_g \mu + \eps \mu_g \ri
+ \xi_g \le
\xi^2_g - \xi^2
- \eps \mu - \eps_g \mu_g \ri \Big],\\
a_0 & = &
 \le \eps_z \mu_z  - \xi^2_z \ri \Big[ \le \eps^2 - \eps^2_g \ri \le
\mu^2 - \mu^2_g \ri + \le \xi^2_g - \xi^2 \ri^2
- 2  \le \xi^2_g + \xi^2 \ri \le \eps \mu + \eps_g \mu_g \ri
\nonumber \\ &&
+ 4 \xi \xi_g \le \eps \mu_g + \mu \eps_g \ri \Big].
\end{eqnarray}
Hence, four relative wavenumbers $\tilde{k} = \kappa_i$, $\kappa_{ii}$,
$\kappa_{iii}$ and $\kappa_{iv}$ may be  extracted~---~either
algebraically or numerically \c{AS}~---~as  the roots of \r{quartic}.

Upon substituting $\hat{\#u} = \hat{\#x}\, \sin \theta  +  \hat{\#z}\,\cos
\theta   $
into  \r{P_pw} and combining with \r{FCM}, the component of $\#P (\#r)$ aligned with $\hat{\#u}$
emerges  straightforwardly  as
\begin{eqnarray}
\hat{\#u} \. \#P (\#r) &=&
\frac{1}{2 \etao}  \exp \le - 2 \ko \tilde{k}_I \,  \hat{\#u}\.\#r \ri
 \,
 \mbox{Re} \, \Bigg\{
 \, \frac{1}{\mu^*_z} \le \tilde{k}^* \sin \theta | E_{0y}|^2 -
i \xi^*_z  E_{0y} E^*_{0z} \ri \, \sin \theta
\nonumber \\ &&
+ \frac{1}{(\mu^*)^2 - (\mu^*_g)^2} \, \Bigg[ \tilde{k}^* \Bigg(
  \les
\mu^* \le |E_{0x}|^2 + |E_{0y}|^2 \ri
+
i \mu^*_g \le E_{0x} E^*_{0y} - E_{0y} E^*_{0x} \ri \ris \,
\cos^2 \theta \nonumber \\ &&  +
\mu^* |  E_{0z} |^2  \sin^2 \theta
- \les \mu^* \le E_{0z} E^*_{0x} + E_{0x}
E^*_{0z} \ri + i \mu^*_g \le E_{0z}E^*_{0y} - E_{0y} E^*_{0z} \ri \ris
\,  \sin \theta \cos \theta \,
\Bigg)
\nonumber \\ &&
+ \le \mu^* \xi^*_g - \mu^*_g \xi^* \ri \les \le
|E_{0x}|^2 +  |E_{0y}|^2 \ri \, \cos \theta - E_{0z} E^*_{0x} \sin \theta \ris
\nonumber \\ &&
- i \le \mu^* \xi^* - \mu^*_g \xi^*_g \ri \les
\le E_{0x} E^*_{0y} - E^*_{0x} E_{0y}
\ri \, \cos \theta  - E_{0z} E^*_{0y}  \sin \theta \ris \Bigg] \Bigg\}, \l{kP_z}
\end{eqnarray}
wherein $(E_{0x},  E_{0y}, E_{0z}) = \#E_0$.

Let the quantity
\begin{equation}
w =
2 \etao
 \exp \le  2 \ko \tilde{k}_I \,   \hat{\#u}\.\#r\, \ri
 |E_{0y}|^{-2} \,  \tilde{k}_R \, \hat{\#u} \. \#P (\#r)\,
 \l{wz}
\end{equation}
be introduced
such that  the fulfilment of the negative phase--velocity condition
\begin{equation}
   \tilde{k}_R \, \hat{\#u} \. \#P (\#r) < 0
   \end{equation}
   is signaled by  $w < 0$.

Substitution of \r{kP_z} in \r{wz} yields the expression
\begin{eqnarray}
w &=& \tilde{k}_R \,
 \mbox{Re} \, \Bigg\{
 \,\frac{1}{\mu^*_z}
 \le \tilde{k}^* \sin \theta  -
i \xi^*_z  \beta^* \ri \,\sin \theta
\nonumber \\ &&
+ \frac{1}{(\mu^*)^2 - (\mu^*_g)^2} \, \Bigg[ \tilde{k}^* \Bigg(
 \les
\mu^* \le |\alpha|^2 +1 \ri
+
i \mu^*_g \le \alpha - \alpha^* \ri \ris \, \cos^2 \theta  \nonumber \\ &&  +
\mu^* | \beta |^2  \sin^2 \theta
- \les \mu^* \le \alpha^* \beta +  \alpha \beta^*
\ri
+ i \mu^*_g \le \beta - \beta^* \ri \ris \, \sin \theta \cos \theta
\Bigg)
\nonumber \\ &&
+ \le \mu^* \xi^*_g - \mu^*_g \xi^* \ri \les \le
|\alpha|^2 +  1 \ri \,  \cos \theta - \alpha^* \beta \,  \sin \theta \ris
\nonumber \\ &&
- i \le \mu^* \xi^* - \mu^*_g \xi^*_g \ri \les
\le \alpha - \alpha^*
\ri\,  \cos \theta  - \beta \,  \sin \theta \ris \Bigg] \Bigg\}
 \,. \l{w_xz}
\end{eqnarray}
The ratios of electric field components
\begin{equation}
\left.
\begin{array}{l}
\alpha = E_{0x} / E_{0y} \\[5pt]
\beta = E_{0z} / E_{0y}
\end{array}
\right\}
\end{equation}
in \r{w_xz} are  derived as follows:
As a function of $\theta$, the  dyadic operator $\=L$  of  \r{L_nabla}
has the form
\begin{eqnarray}
 \=L &=&
i \ko \Big\{
\les \,  \=L \, \ris_{11}
 \hat{\#x}\, \hat{\#x} +
\les \,  \=L \, \ris_{22}
 \hat{\#y}\, \hat{\#y}
+ \les \,  \=L \, \ris_{33}
 \hat{\#z}\, \hat{\#z}
+\les \,  \=L \, \ris_{12}
\le \,  \hat{\#x}\, \hat{\#y}
-  \hat{\#y}\, \hat{\#x} \, \ri \nonumber \\ &&
+\les \,  \=L \, \ris_{13}
\le \,  \hat{\#x}\, \hat{\#z}
+  \hat{\#z}\, \hat{\#x} \, \ri
+\les \,  \=L \, \ris_{23}
\le \,  \hat{\#y}\, \hat{\#z}
-  \hat{\#z}\, \hat{\#y} \, \ri \Big\},
\end{eqnarray}
with components
\begin{eqnarray}
\les \,  \=L \, \ris_{11} \l{L_11}
&=&
\eps + \frac{2 \mu_g \xi \Gamma - \mu \le \xi^2 + \Gamma^2 \ri}{\mu^2
- \mu^2_g},
\\
\les \,  \=L \, \ris_{22}
&=&
\eps - \frac{\tilde{k}^2 \sin^2 \theta}{\mu_z} + \frac{ 2 \mu_g   \xi \Gamma - \mu \le \xi^2 + \Gamma^2
\ri }{\mu^2
- \mu^2_g},\\
\les \,  \=L \, \ris_{33}
&=& \eps_z - \frac{\xi^2_z}{\mu_z} - \frac{\mu \tilde{k}^2 \sin^2 \theta}{\mu^2
- \mu^2_g},\\
\les \,  \=L \, \ris_{12} \l{L_12}
&=& i \le
\eps_g + \frac{\mu_g \le \xi^2 + \Gamma^2 \ri - 2 \mu \xi \Gamma}{\mu^2
- \mu^2_g} \ri,\\
\les \,  \=L \, \ris_{13}
&=&  \frac{  \mu \Gamma - \mu_g \xi }{\mu^2
- \mu^2_g} \, \tilde{k} \, \sin \theta \, ,\\
\les \,  \=L \, \ris_{23}
&=&  i  \le
\frac{\mu_g \Gamma -  \mu \xi  }{\mu^2
- \mu^2_g}- \frac{\xi_z}{\mu_z} \ri  \, \tilde{k} \, \sin \theta \,,
\end{eqnarray}
where $\Gamma = \xi_g + \tilde{k} \, \cos \theta $.
It follows  from the vector Helmholtz equation
 \r{Helm} that
\begin{equation}
\left.
\begin{array}{l}
\alpha = \displaystyle{ \frac{  \les \,  \=L \, \ris_{12}  \les \,  \=L \,
\ris_{33} +    \les \,  \=L \, \ris_{13}
  \les \,  \=L \, \ris_{23} }
{  \les \,  \=L \, \ris_{13}  \les \,  \=L \, \ris_{13}
-   \les \,  \=L \, \ris_{11}  \les \,  \=L \, \ris_{33}}} \\
\vspace{-3mm}
\\
\beta =
\displaystyle{
\frac{   \les \,  \=L \, \ris_{12}  \les \,  \=L \, \ris_{23}  -
  \les \,  \=L \, \ris_{13}  \les \,  \=L \, \ris_{22} }
{  \les \,  \=L \, \ris_{13}  \les \,  \=L \, \ris_{23} +
  \les \,  \=L \, \ris_{12}  \les \,  \=L \, \ris_{33} }}
\end{array}
\right\}. \l{alp_bet}
\end{equation}

\subsection{ Propagation along the  $z$ axis} \label{2.z}

The results of the preceding analysis simplify considerably for planewave
propagation along the $z$ axis (i.e.,  $ \theta =  0$).
 The quartic dispersion relation  \r{quartic}
yields the four relative wavenumbers
\begin{equation}
\left.
\begin{array}{l}
\kappa_{i} =       \sqrt{ \eps  +
 \eps_g  } \sqrt{ \mu +  \mu_g} - \xi - \xi_g   \\
\kappa_{ii} =    -  \sqrt{ \eps  +
 \eps_g  } \sqrt{ \mu +  \mu_g} - \xi - \xi  \\
\kappa_{iii} =  \sqrt{ \eps  -
 \eps_g  } \sqrt{ \mu - \mu_g}
 +  \xi -  \xi_g
\, \\
\kappa_{iv} =  - \sqrt{ \eps  -
 \eps_g  } \sqrt{ \mu -  \mu_g}
+  \xi -  \xi_g
\end{array}
\right\}\,; \l{roots_kz}
\end{equation}
and    \r{w_xz}  reduces to
\begin{eqnarray}
w = \tilde{k}_R \, \mbox{Re} \lec
 \frac{  \le
| \alpha |^2 + 1 \ri \le \,
\tilde{k}^* \mu^*
 - \mu^*_g \xi^* + \mu^* \xi^*_g \ri
+ i
 \le \alpha - \alpha^*  \ri
\le \,
\tilde{k}^*  \mu^*_g -
  \mu^*  \xi^* + \mu^*_g \xi^*_g \ri}
{(\mu^*)^2 - (\mu^*_g)^2} \, \ric  \,. \l{cond_z}
\end{eqnarray}
Since the dyadic operator components $\les\,\=L \,\ris_{13}$ and
 $\les\,\=L \,\ris_{23}$ are null--valued for $\theta = 0$, the electric field ratios
are
given as
\begin{equation}
\left.
\begin{array}{l}
\alpha = -  \les \,  \=L \, \ris_{12} / \les \,
\=L \, \ris_{11}\\
\beta = 0
\end{array}
\right\}.
\end{equation}
Note that a further consequence of  $\les\,\=L \,\ris_{13} =
\les\,\=L \,\ris_{23} = 0$ is that the time--averaged Poynting
vector is parallel to the $z$ axis.

By substituting  \r{roots_kz} into \r{L_11} and \r{L_12}, the ratio
$\alpha$
emerges  as
\begin{equation}
\alpha =
\left\{
\begin{array}{lcccl}
i && \mbox{for} && \tilde{k} = \kappa_{i}, \kappa_{ii}, \\
 -i &&  \mbox{for} &&  \tilde{k} = \kappa_{iii}, \kappa_{iv}.
\end{array}
\right. \l{alpha_val}
\end{equation}
Hence, negative--phase velocity propagation along the $z$ axis
occurs provided $w < 0$ where
\begin{eqnarray}
&&  w = w_i = 2\,  \mbox{Re}\lec
 \sqrt{\eps + \eps_g} \sqrt{\mu + \mu_g}  - \xi -
\xi_g \ric \, \mbox{Re} \lec \frac{ \sqrt{\eps^* + \eps^*_g}}{ \sqrt{\mu^* + \mu^*_g}} \ric
 \qquad \mbox{for} \qquad \tilde{k} = \kappa_{i}, \nonumber \\ && \\
&& w = w_{ii} = 2\, \mbox{Re}\lec
 \sqrt{\eps + \eps_g} \sqrt{\mu + \mu_g}  + \xi +
\xi_g \ric \, \mbox{Re} \lec \frac{ \sqrt{\eps^* + \eps^*_g}}{
\sqrt{\mu^* + \mu^*_g}} \ric
 \qquad \mbox{for} \qquad \tilde{k} = \kappa_{ii}, \nonumber \\ && \\
&& w = w_{iii} =  2\,
 \mbox{Re}\lec
 \sqrt{\eps - \eps_g} \sqrt{\mu - \mu_g}  + \xi -
\xi_g \ric \, \mbox{Re} \lec \frac{ \sqrt{\eps^* - \eps^*_g}}{ \sqrt{\mu^* - \mu^*_g}} \ric
  \qquad \mbox{for} \qquad \tilde{k} = \kappa_{iii}, \nonumber \\ &&
\\
&& w = w_{iv} = 2\, \mbox{Re}\lec
 \sqrt{\eps - \eps_g} \sqrt{\mu - \mu_g}  - \xi +
\xi_g \ric \, \mbox{Re} \lec \frac{ \sqrt{\eps^* - \eps^*_g}}{
\sqrt{\mu^* - \mu^*_g}} \ric
   \qquad \mbox{for} \qquad \tilde{k} = \kappa_{iv}. \nonumber \\
\end{eqnarray}

\section{Numerical results}

In order to further examine the negative phase--velocity conditions
derived in Sections \ref{2.xz} and \ref{2.z}, let us consider a 
Faraday chiral medium
(FCM)
produced by mixing
  (a) an isotropic chiral medium described by the
constitutive relations \c{Beltrami}
\begin{equation}
\left.
\begin{array}{l}
  \#D = \epso \eps^a \, \#E + i \sqrt{\epso \muo}\, \xi^a \, \#H \,\\[5pt]
  \#B = -  i \sqrt{\epso \muo} \, \xi^a \, \#E + \muo \mu^a \,\#H \,
\l{chiral}
\end{array}
\right\}
\end{equation}
and (b)  a magnetically biased ferrite
  described by the
constitutive relations \cite[Ch. 7]{Chen}
\begin{equation}
\left.
\begin{array}{l}
  \#D = \epso \eps^b \, \#E \\[5pt]
  \#B =
\muo \les \, \mu^b \, \=I
- i \mu^b_g \, \hat{\#z} \times \=I +
\le \, \mu^b_z - \mu^b \, \ri \,  \hat{\#z}\,  \hat{\#z} \, \ris
  \.\#H \,
\l{mag}
\end{array}
\right\}.
\end{equation}
Both  component mediums are envisioned as random distributions of
electrically small, spherical particles.
The resulting homogenized composite medium (HCM) is a FCM  characterized by
the
  constitutive dyadics
\begin{equation}
\left.
\begin{array}{l}
  \=\eps^{HCM} =
\epso \les \, \eps^{HCM} \, \=I
- i \eps^{HCM}_g \, \hat{\#z} \times \=I +
\le \, \eps^{HCM}_z - \eps^{HCM} \, \ri \,  \hat{\#z}\,  \hat{\#z} \, \ris\\
\vspace{-2mm} \\
  \=\xi^{HCM} =
i \sqrt{\epso \muo} \, \les \, \xi^{HCM} \, \=I
- i \xi^{HCM}_g \, \hat{\#z} \times \=I +
\le \, \xi^{HCM}_z - \xi^{HCM} \, \ri \,
  \hat{\#z} \,  \hat{\#z}
\ris\\
\vspace{-2mm} \\
  \=\mu^{HCM} =
\muo \les \, \mu^{HCM} \, \=I
- i \mu^{HCM}_g \, \hat{\#z} \times \=I +
\le \, \mu^{HCM}_z - \mu^{HCM} \, \ri \,  \hat{\#z}\,  \hat{\#z} \, \ris
\end{array}
\right\}. \l{HCM}
\end{equation}
Incidentally, a FCM with constitutive dyadics of the
form \r{HCM} may also be developed via the homogenization of an
isotropic chiral medium and a
magnetically biased plasma \c{WM00}.

The constitutive
dyadics $ \=\eps^{HCM}$, $ \=\xi^{HCM}$ and $ \=\mu^{HCM}$ are
estimated using the Bruggeman homogenization formalism for a
representative
  example.
Comprehensive details of the Bruggeman formalism \c{Ward, M03} and
its implementation
in the context of FCMs \c{WLM98, WM00, MLWM01} are available elsewhere.
Initially, we restrict our attention  to nondissipative FCMs; the influence
of dissipation is considered later in this section.

\subsection{Nondissipative FCMs}
The  parameter values  selected  for nondissipative
component mediums are as follows:
\begin{eqnarray*}
&& \eps^a = 3.2, \:\: \xi^a = 2.4, \:\: \mu^a = 2; \:\: \eps^b = 2.2, \:\:
\mu^b = 3.5, \:\: \mu^b_z = 1, \:\: \mu^b_g \in [0,4].
\end{eqnarray*}
The permeability parameters for component medium $b$
may be viewed in terms
of the semi--classical ferrite model as
\begin{equation}
\left.
\begin{array}{l}
\mu^b = \displaystyle{1 + \frac{\omegao \, \omega_m}{\omegao^2 -
\omega^2}}\\[8pt]
\mu^b_g =  \displaystyle{ \frac{\omega \, \omega_m}{\omegao^2 -
\omega^2}}\\[8pt]
\mu^b_z = 1
\end{array}
\right\},
\end{equation}
wherein $\omegao$ is the Larmor precessional frequency
of spinning electrons
 and $\omega_m$ is the saturated magnetization frequency
\c{Lax, Chen}.
Thus, the parameter values $\mu^b = 3.5$ and $\mu^b_g \in [0,4]$
correspond to the relative frequency range
$\le \omegao / \omega \ri \in [0.625, \infty )$.

Let $f_a$ denote the volume fraction of the isotropic chiral component
medium $a$.
In figure 1, the estimated constitutive parameters of the HCM are plotted as
functions of $f_a$ for $\mu^b_g = 4$.
The uniaxial and gyrotropic characteristics of a FCM
are clearly reflected by the constituents of the
permeability dyadic $\=\mu^{HCM}$ and the magnetoelectric dyadic
$\=\xi^{HCM}$. In contrast,
the
HCM is  close to being isotropic with respect to its
dielectric properties.
Significantly, eight of the nine scalars appearing in \r{HCM} are
positive, while
$\eps_g^{HCM}$ is negative only for $f_a < 0.32$; however,
$\vert\eps_g^{HCM}\vert << \vert\eps^{HCM}\vert$ and
$\vert\eps_g^{HCM}\vert << \vert\eps_z^{HCM}\vert$ for all values of
$f_a \in [0,1]$.

 The permeability parameters  $\mu^{HCM}$ and
$\mu^{HCM}_g$ are equal
at $f_a \approx 0.25$, it being clear from the right side
of \r{w_xz}  that
this equality
has an important bearing on the stability of $w$.
Further
calculations with $\mu^b_g = 2$ and $\mu^b_g = 3$ have
  confirmed that  $\mu^{HCM} \neq
\mu^{HCM}_g$ for all volume fractions  $f_a \in [0,1]$.
This matter is pursued in figure 2 where
the estimated constitutive parameters of the HCM are graphed as
functions of  $\mu^b_g$
for $f_a = 0.35$. The HCM gyrotropic parameters
$\xi^{HCM}_g$ and $\mu^{HCM}_g$ are  observed to increase steadily as
  $\mu^b_g$ increases;  $\eps^{HCM}_g$,
$\xi^{HCM}_g$ and $\mu^{HCM}_g$ all
 vanish in the
limit $\mu^b_g \rightarrow 0$. Also, as $\mu^b_g$ increases,
the degree of uniaxiality (with respect to the $z$ axis) increases for
$\=\xi^{HCM}$ but
  decreases for   $\=\mu^{HCM}$.

The relative wavenumbers $\tilde{k} = \kappa_{i-iv}$ for propagation along the $z$
axis, as specified in
 \r{roots_kz},
 are displayed in figure 3
as  functions of  $f_a$, for $\mu^b_g = 2, 3$ and $4$.
The relative wavenumbers $\kappa_i > 0 $ and $\kappa_{ii} < 0$
 for all $f_a \in [0,1]$ for $\mu^b_g = 2, 3$ and $4$.
 Similarly, for $\mu^b_g = 2$ and $3$,  the
relative numbers  $\kappa_{iii} > 0$ and $\kappa_{iv} < 0$.

However, the
equality  $\mu^{HCM} =
\mu^{HCM}_g$,  which occurs at $f_a \approx 0.25$ for
$\mu^b_g = 4$,  results in
   $\kappa_{iii}$ and $\kappa_{iv}$ acquiring
 nonzero imaginary
 parts as $f_a$ falls below $0.25$ for  $\mu^b_g = 4$.
Only the real parts of these complex--valued relative
wavenumbers are plotted
in figure 3.

 Observe that
  $\kappa_i$, $\kappa_{iii}$ and  $\kappa_{iv}  > 0$ in figure 3,  whereas
$\kappa_{ii} < 0$ in the volume fraction range
  $ 0.25 < f_a < 0.42$ with  $\mu^b_g = 4$. Furthermore,
 $\kappa_i$, $\mbox{Re} \, \lec
\kappa_{iii} \, \ric$ and  $\mbox{Re} \, \lec
\kappa_{iv} \, \ric > 0$ while $\kappa_{ii} < 0$  for $ f_a < 0.25$ with  $\mu^b_g = 4$.
In the limit $f_a \rightarrow 0$, the relative wavenumbers
$ \kappa_{i-iv} \rightarrow \pm \sqrt{\eps^b}\sqrt{\mu^b
\pm \mu^b_g}$ (i.e.,
the relative wavenumbers of a ferrite biased along the $z$ axis
\c{Chen}). Also, as
  $f_a \rightarrow 1$, the relative wavenumbers
$ \kappa_{i-iv} \rightarrow \pm \sqrt{\eps^a
\mu^a} \pm \xi^a$ (i.e., the
relative wavenumbers of an isotropic chiral medium).

The
values of $w$ corresponding to the relative wavenumbers
$\kappa_{i-iv}$ of figure 3, namely $w_{i-iv} $, are
plotted against $f_a$ in figure 4  for $\mu^b_g = 2, 3$ and $4$.
The quantities
$w_{i-iii} \ge 0$
for all volume fractions $f_a \in [0,1]$ with  $\mu^b_g = 2, 3$ and
$4$.
Thus, for the relative wavenumbers $\kappa_{i-iii}$, 
power  flows in the same direction as
the phase velocity. This is the case regardless of whether the phase
velocity is directed along the positive $z$ axis (as in  modes 
 $\kappa_{i}$ and $\kappa_{iii}$) or directed
along the negative  $z$ axis (as in  mode  $\kappa_{ii}$).
Both $w_{iii}$ and $w_{iv}$ are null valued in those regions where
the corresponding relative wavenumbers, $\kappa_{iii}$ and
$\kappa_{iv}$, respectively, have nonzero imaginary parts.
In addition,  $w_{iii} \rightarrow \infty$ and $w_{iv} \rightarrow - \infty$
  in the vicinity of $f_a = 0.25$ for $\mu^b_g = 4$.

Significantly,
  $w_{iv} <  0$   for $\mu^b_g = 4$ at volume
fractions $f_a \in (0.25, 0.42)$ in figure 4. This means that {\em  the negative
phase--velocity
condition then holds in the chosen FCM\/} 
which has been conceptualized as a homogenized composite medium.

In figure 5, the relative wavenumbers $\kappa_i$ and $\kappa_{iii}$
  for  $\theta = \pi /2$ (i.e., propagation along the $x$ axis) are
plotted against   $f_a$ for $\mu^b_g = 2, 3$ and $4$.
The
graphs of  $\kappa_{ii}$ and  $\kappa_{iv}$ need not be presented
since
  $\kappa_{i} = - \kappa_{ii}$ and  $\kappa_{iii} = -
\kappa_{iv}$.\footnote{When $\theta = \pi/2$, the
dispersion relation \r{quartic}
reduces to a  quadratic polynomial in $\tilde{k}^2$.}
For all $f_a \in [0,1]$ with $\mu^b_g = 2$ and $3$, the
 relative wavenumbers $\kappa_{i} > 0$ and $\kappa_{iii} < 0$.
Similarly, $\kappa_i > 0$  for $\mu^b_g = 4$.
However,
when
 $\mu^b_g = 4$, it is found  that $\kappa_{iii} < 0$ for $f_a >
0.42$
but $\kappa_{iii}$ possesses a nonzero imaginary part and
$\mbox{Re} \, \lec \kappa_{iii} \ric = 0 $ for $f_a < 0.42$.
In the limit $f_a \rightarrow 0$, the relative wavenumbers
$\kappa_{i-iv} \rightarrow \pm \sqrt{\eps^b / \mu^b
}\sqrt{\le \mu^b \ri^2 -
\le \mu^b_g \ri^2}$ and  $\pm\sqrt{\eps^b \mu^b_z}$
  (i.e.,
the relative wavenumbers of a  ferrite biased
along the $x$ axis \c{Chen}). Also, as
  $f_a \rightarrow 1$, the relative wavenumbers
$\kappa_{i-iv} \rightarrow \pm \sqrt{\eps^a
\mu^a} \pm \xi^a$ (i.e., the
relative wavenumbers of an isotropic chiral medium).

Figure 6 shows the plots
  of $w_{i,iii} $ corresponding to
the relative wavenumbers
$\kappa_{i,iii}$ of figure 5.
The graphs of  $w = w_{ii,iv} $ corresponding to the relative
waveumbers $\kappa_{ii,iv}$ are not displayed since
 the equalities $w_i = w_{ii}$ and $w_{iii} = w_{iv}$ hold
for $\theta = \pi/2$~---~as may be inferred from \r{w_xz}--\r{alp_bet}.
The quantities $w_{i,iii} \ge 0$ at
all volume fractions $f_a \in [0,1]$ with  $\mu^b_g = 2, 3$ and $4$.
As remarked earlier for 
  $\kappa_{i-iii}$
propagation along the $z$ axis, here we have that power flows in the
same direction as the phase velocity, regardless of whether
the phase velocity is directed along the positive $x$ axis (mode
$\kappa_{i}$)
or along the negative $x$ axis (mode  $\kappa_{iii}$).
Furthermore, it is found that
$w_{iii} = 0$ in the region where the corresponding relative
wavenumber $\kappa_{iii}$ is purely imaginary (i.e., for $f_a < 0.42$
with $\mu^b_g = 4$).

\subsection{Dissipative FCMs}
The scope of these numerical investigations
 is now broadened
by considering (i) the effects of dissipation or loss; and (ii)
propagation in an arbitrary direction.
Let a small amount of loss be incorporated into
 component medium  $b$ by selecting the constitutive parameters
of the component mediums as
\begin{eqnarray*}
&& \eps^a = 3.2, \:\: \xi^a = 2.4, \:\: \mu^a = 2; \:\: \eps^b = 2.2 +
i \, \delta , \:\:
\mu^b = 3.5 + i \, \delta , \:\: \mu^b_z = 1 +  i \, 0.5 \delta , \:\:
\mu^b_g = 4 + i \, 2 \delta,
\end{eqnarray*}
where the dissipation parameter $\delta \in [0,0.2]$.
We focus attention  on
the region of negative phase--velocity
  propagation along the $z$ axis
with  relative wavenumber $\kappa_{iv}$, as
illustrated by $w_{iv} < 0$ at $0.25 < f_a  < 0.42 $
in figure 4.

Real parts of the relative wavenumber $\kappa_{iv}$,
calculated at the volume fraction $f_a = 0.35$ with $\delta = 0, 0.1$
and $0.2$,
are  graphed as  functions of $\theta$ in figure 7.
The relative wavenumber $\kappa_{iv}$ for the nondissipative FCM (i.e.,
$\delta = 0$) is real--valued for $\theta < 52^\circ$ but has a nonzero
imaginary part for $\theta > 52^\circ$. The relative wavenumbers
$\kappa_{iv}$
for $\delta = 0.1$ and $0.2$ have nonzero imaginary parts
for all values of $\theta$. Note that the real part of $\kappa_{iv}$
falls to zero at $\theta = \pi/2$ in the absence of dissipation (i.e., $\delta = 0$).

Plots of the quantity $w = w_{iv}$, corresponding to the relative wavenumber
$\kappa_{iv}$ of figure 7, are provided  in figure 8. {\em The negative
phase--velocity condition
 $w_{iv} < 0$ is satisfied 
 \begin{itemize}
 \item[(i)]
for $\theta < 52^\circ$ when $\delta = 0$,
\item[(ii)] for $\theta < 76^\circ$
when $\delta =0.1$, and 
\item[(iii)] for $\theta < 38^\circ$ when $\delta = 0.2$.
\end{itemize}
}

\section{Discussion and Conclusion}

In isotropic dielectric--magnetic mediums, plane waves can propagate
with  phase velocity directed opposite
 to the direction of
power flow under certain, rather restrictive, conditions
\c{MartinLW02}.
However,
the constitutive parameter space associated with anisotropic and
  bianisotropic mediums provides a wealth of opportunities for
observing and exploiting
negative phase--velocity behavior.
General conditions are established here for the phase velocity to be
directed opposite to power flow for a particular class
of bianisotropic mediums, namely Faraday chiral mediums.
The theory has been explored by means of a
representative example of FCMs, arising from
the homogenization of an isotropic chiral medium and a
magnetically biased ferrite.
For our representative example, the negative phase--velocity conditions
have been found to hold for propagation
in  arbitrary directions~---~for both nondissipative and dissipative
FCMs~---~provided that the gyrotropic parameter
of
the ferrite component medium is sufficiently large compared with the corresponding
nongyrotropic permeability parameters.

Previous studies \c{Ziolkowski}--\c{Hu} have emphasized the importance of
the  signs
 of  constitutive (scalar) parameters in establishing the conditions for
negative phase--velocity propagation in homogeneous mediums.\footnote{Parenthetically,
negative
refraction is also displayed by certain purely dielectric mediums, but
they must be nonhomogeneous \c{Notomi,LJJP}.}
In the absence of dissipation,
 negative phase--velocity propagation has been predicted in 
  \begin{itemize}
 \item[(i)]  isotropic
dielectric--magnetic mediums
provided that both the permittivity and permeability
scalars are negative \c{Veselago03}, and  
\item[(ii)] uniaxial
dielectric--magnetic
mediums when only one of the four
 constitutive scalars
is negative \c{Lindell01}.
\end{itemize}
Also, the conditions
for negative phase--velocity  propagation
may be fulfilled by  dissipative isotropic dielectric--magnetic
mediums when  only
 one of the two constitutive scalars has
a negative  real part  \c{MartinLW02}.
The present study demonstrates that {\em the condition
for  negative phase--velocity
propagation can be satisfied by nondissipative FCMs
with
 constitutive
scalars that are all positive\/}. Furthermore, these  conditions
continue to be satisfied after the introduction a small amount of
dissipation.

For the particular case of propagation  parallel
to the ferrite biasing field, the components
of the time--averaged Poynting vector are null--valued in directions
perpendicular to the propagation direction.
In contrast, for general propagation directions, the 
time--averaged
Poynting vector
has nonzero components perpendicular to the direction of
propagation. Further studies are required to explore
the consequences
of the
 negative phase--velocity condition
$   \tilde{k}_R \, \hat{\#u} \. \#P (\#r) < 0 $
 for such general propagation directions.

To conclude, more general
bianisotropic mediums, particularly those developed as HCMs
based on nonspherical particulate components, offer exciting
prospects for future studies of negative phase--velocity propagation.

\newpage

\begin{figure}[!ht]
\centering \psfull \epsfig{file=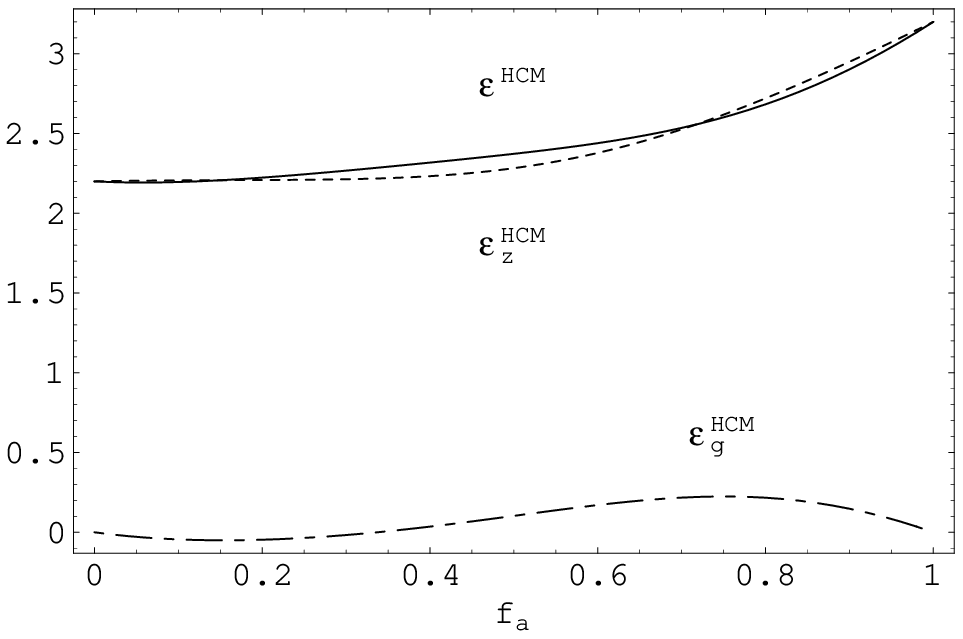,width=3.8in}
\epsfig{file=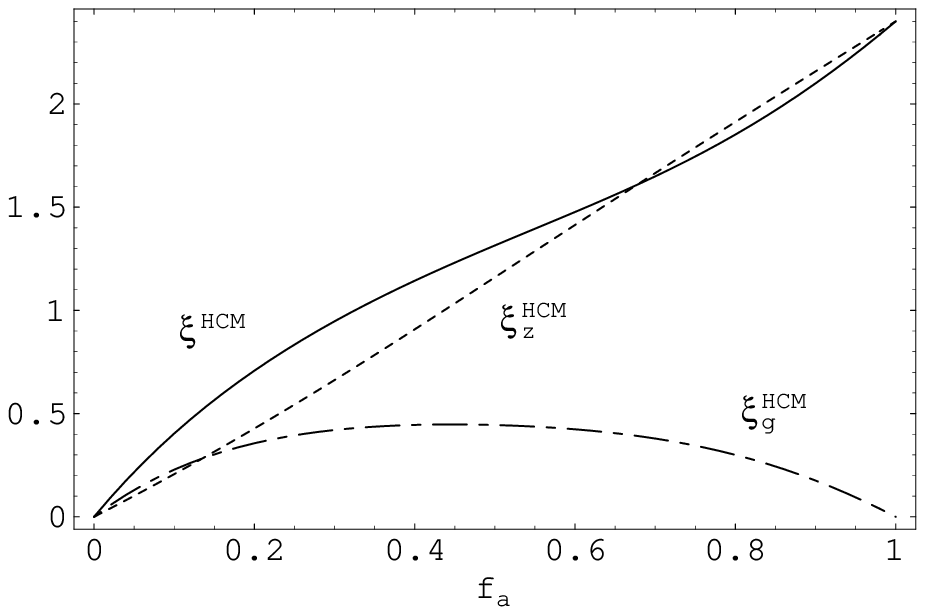,width=3.8in}
\epsfig{file=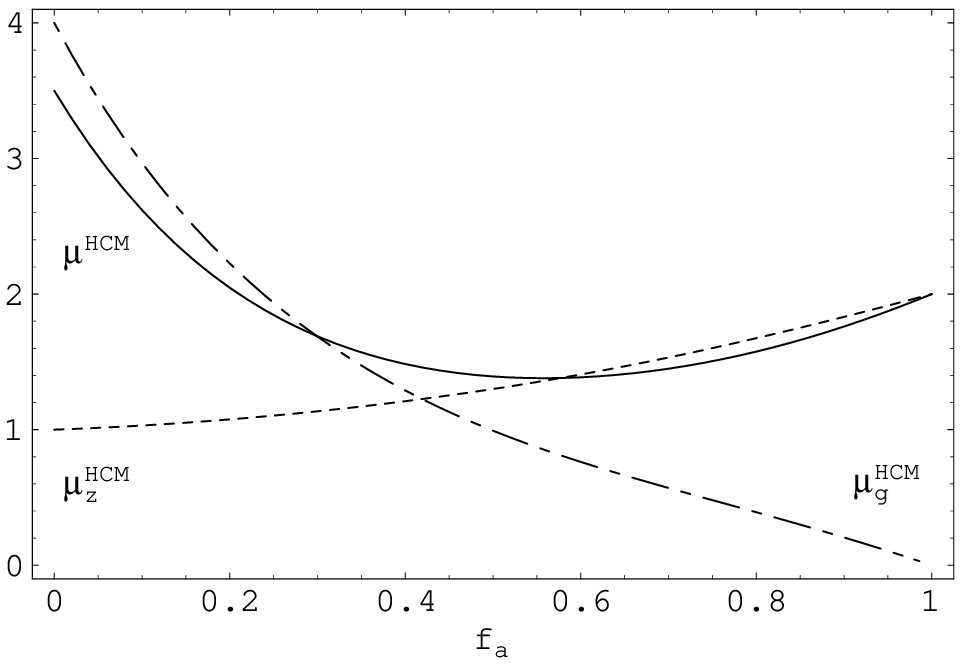,width=3.8in}
  \caption{\label{fig1}
Bruggeman estimates of
$\underline{\underline{\epsilon}}^{HCM}$,
$\underline{\underline{\xi}}^{HCM}$
and $\underline{\underline{\mu}}^{HCM}$ as  functions of $f_a$,
when  $\epsilon^a = 3.2$, $ \xi^a = 2.4$ , $ \mu^a = 2$ ,
$\epsilon^b = 2.2$,
$\mu^b = 3.5$, $\mu^b_z = 1$, and  $ \mu^b_g = 4$.}
\end{figure}

\newpage

\begin{figure}[!ht]
\centering \psfull \epsfig{file=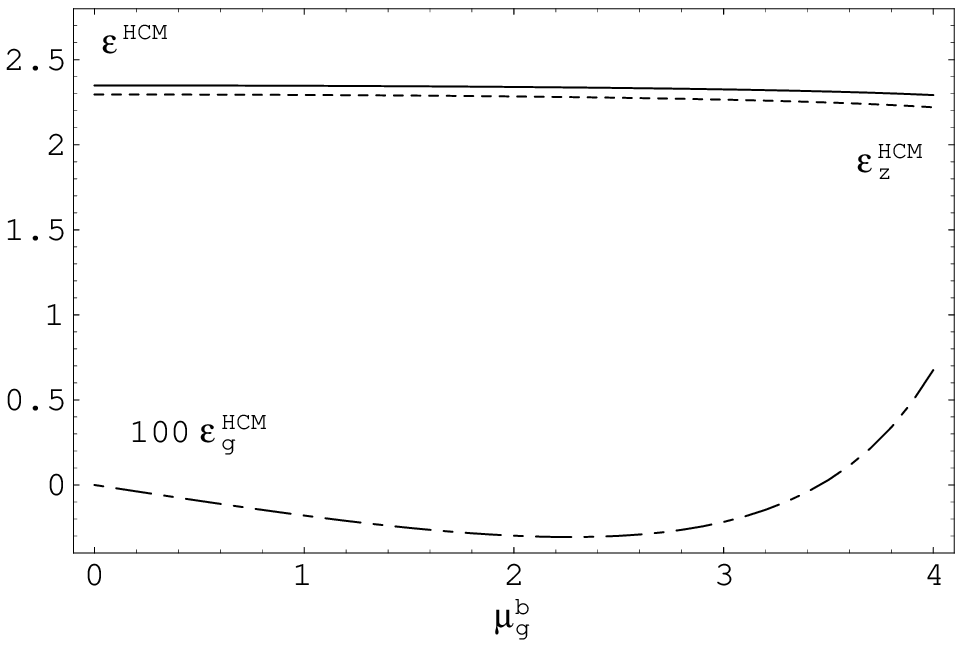,width=3.8in}
\epsfig{file=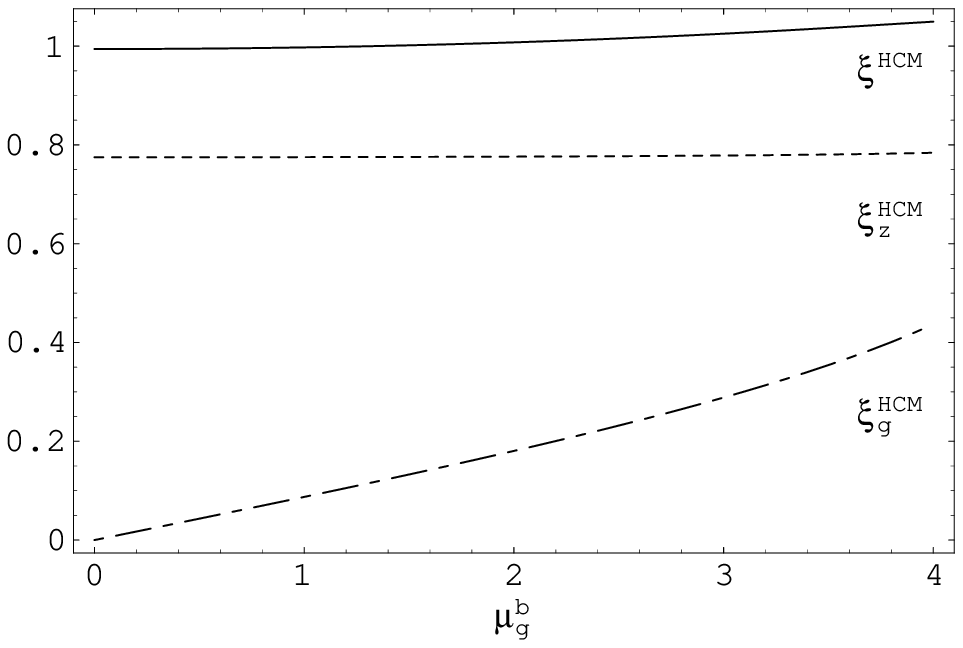,width=3.8in}
\epsfig{file=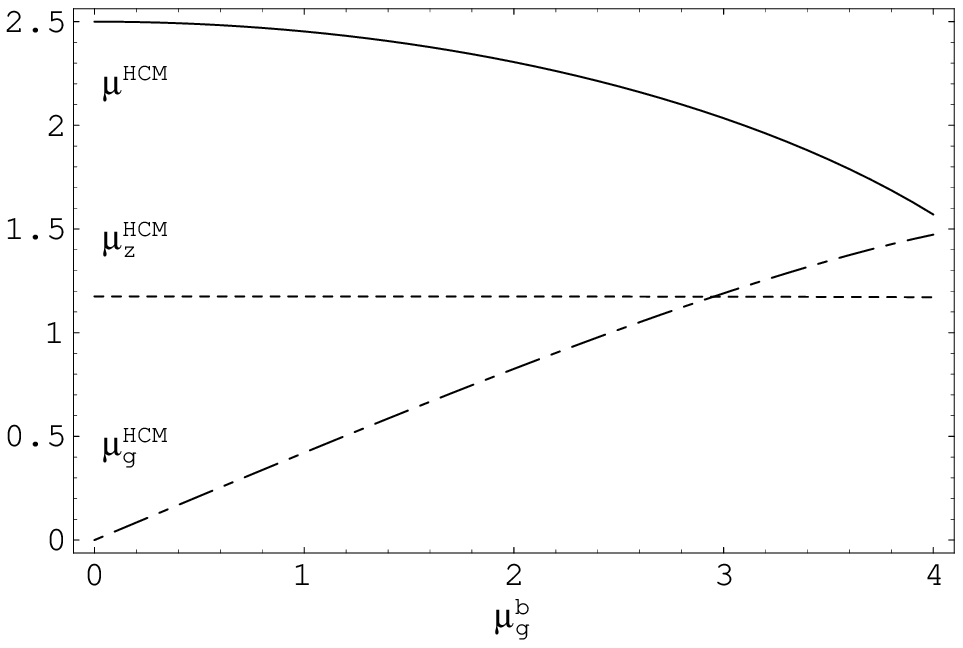,width=3.8in}
  \caption{\label{fig2}
Bruggeman estimates of $\underline{\underline{\epsilon}}^{HCM}$,
$\underline{\underline{\xi}}^{HCM}$
and $\underline{\underline{\mu}}^{HCM}$ as  functions of $\mu^b_g$.
The constitutive parameters of the
component mediums are the same
 as in figure 1, but with $\mu^b_g \in [0,4]$ and $f_a = 0.35$.}
\end{figure}

\newpage

\begin{figure}[!ht]
\centering \psfull \epsfig{file=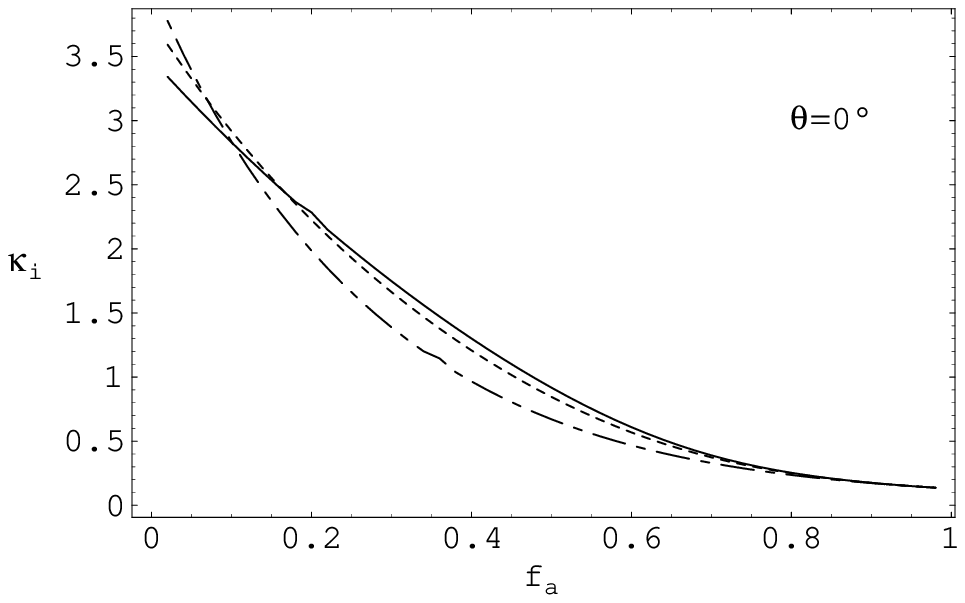,width=3.2in} \hfill
  \epsfig{file=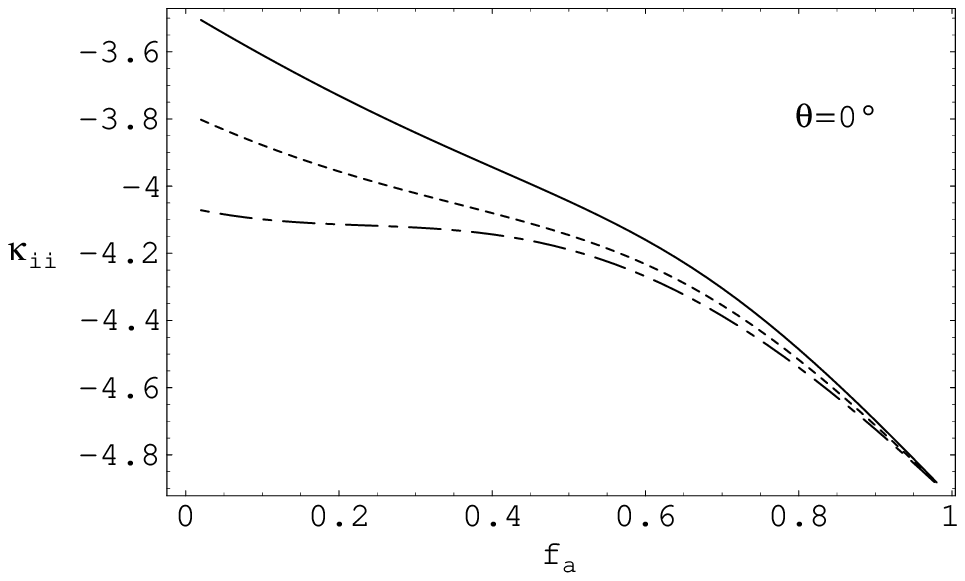,width=3.2in}\\
\epsfig{file=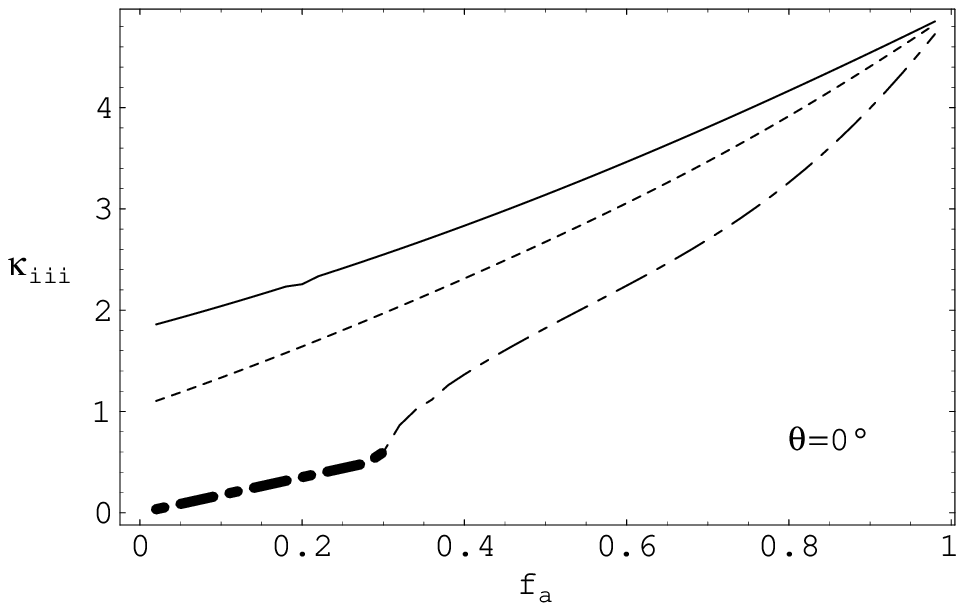,width=3.2in} \hfill
\epsfig{file=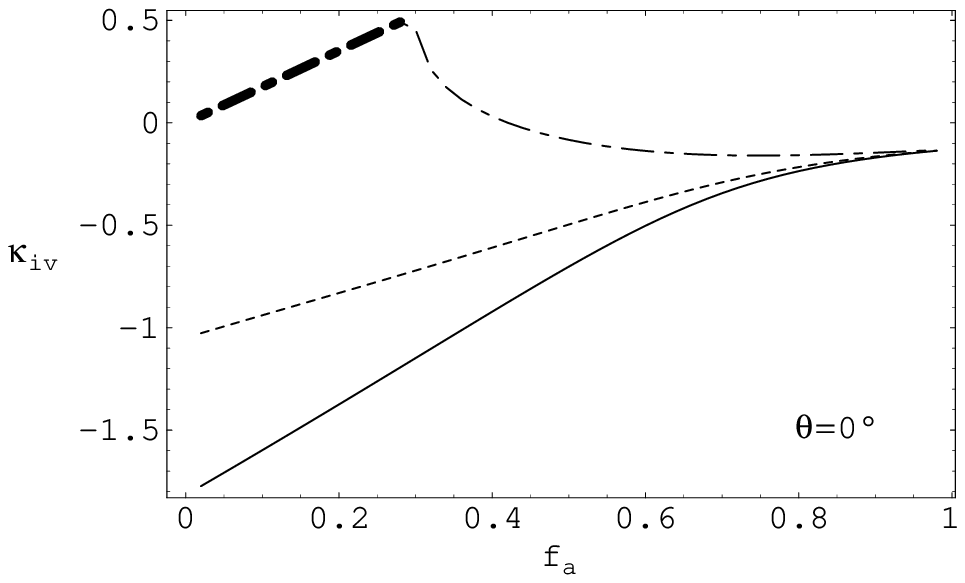,width=3.2in}
  \caption{\label{fig3}
Calculated values of the relative wavenumbers
$\kappa_{i-iv}$ as  functions of $f_a$,  when $\theta=0$ and  $\mu^b_g = 2,
3$ and $4$. 
The constitutive parameters of the
component mediums are:
  $\epsilon^a = 3.2$, $ \xi^a = 2.4$, $ \mu^a = 2$,  $
\epsilon^b = 2.2$,
$\mu^b = 3.5$,  and  $ \mu^b_z = 1$.
  Key:  $\kappa_{i-iv}$ values corresponding to $\mu^b_g =
2, 3$ and $4$ are represented
by the solid lines, dashed lines,
and broken dashed  lines, respectively.
Heavy lines
 indicate those relative wavenumbers
which have nonzero imaginary parts; the real parts
of   such
complex--valued
relative wavenumbers are plotted.
}
\end{figure}

\newpage

\begin{figure}[!ht]
\centering \psfull \epsfig{file=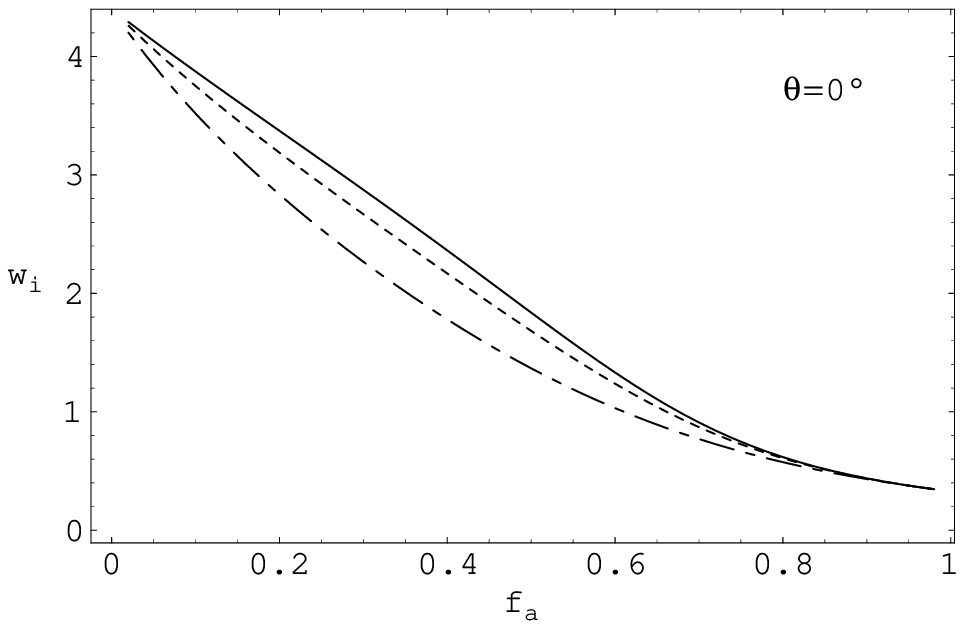,width=3.2in} \hfill
  \epsfig{file=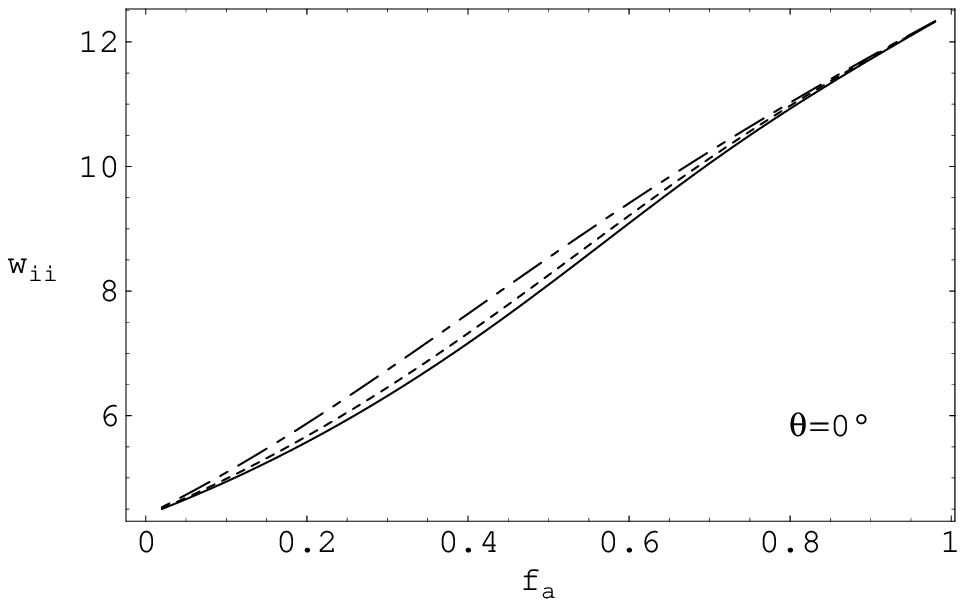,width=3.2in}\\
\epsfig{file=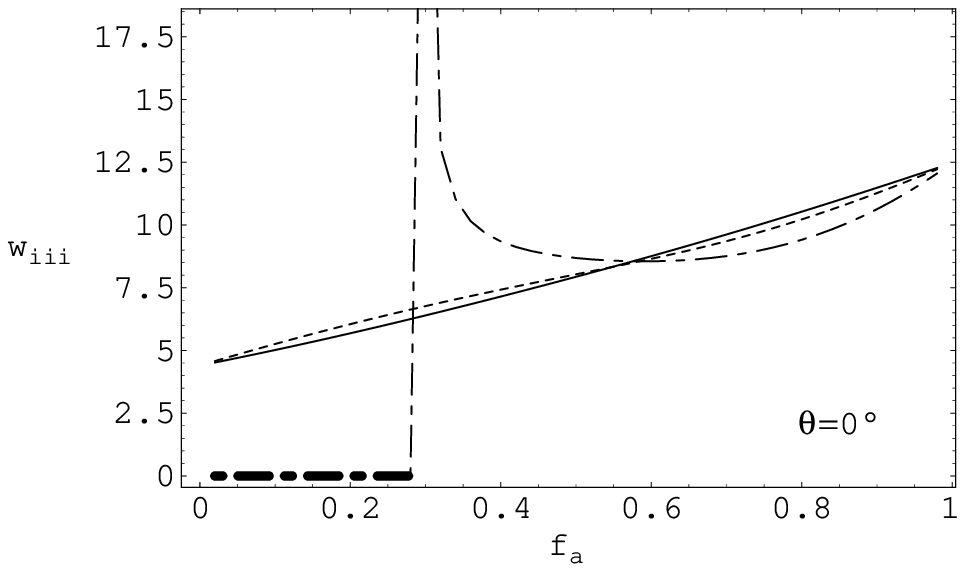,width=3.2in} \hfill
\epsfig{file=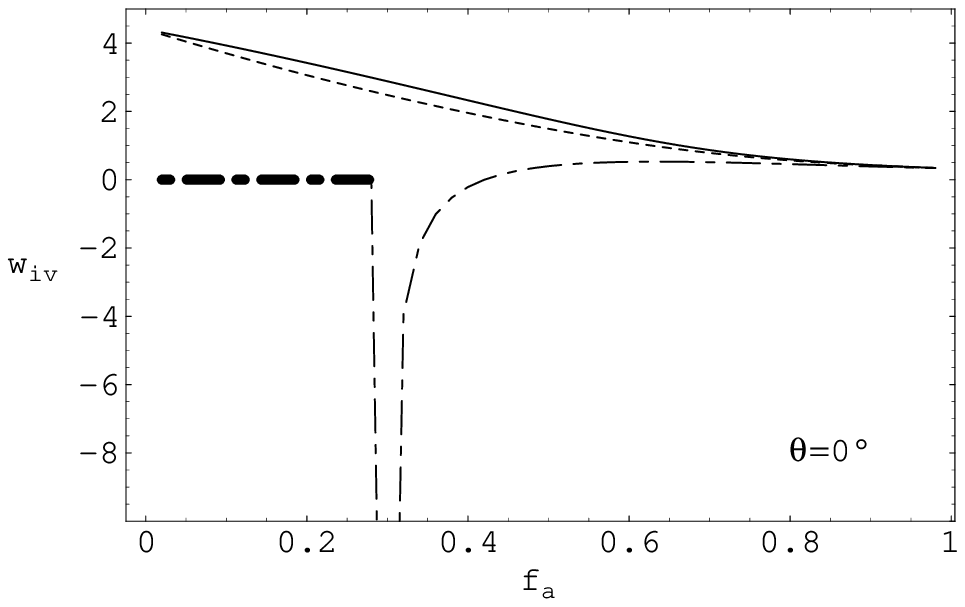,width=3.2in}
  \caption{\label{fig4}
Calculated values of 
$w_{i-iv}$
 as  functions of $f_a$,   when $\theta=0$ and $\mu^b_g = 2, 3$ and $4$.
  The constitutive parameters of the
component mediums are the same as in figure \ref{fig3}.
  Key:  $w_{i-iv}$ values corresponding to $\mu^b_g = 2,
3$ and $4$ are represented
by the solid lines, dashed lines,
and broken dashed  lines, respectively.
Heavy lines
 indicate
those $w$
values that devolve from relative
wavenumbers $\tilde{k}$ with
 nonzero imaginary parts.
}
\end{figure}

\newpage

\begin{figure}[!ht]
\centering \psfull \epsfig{file=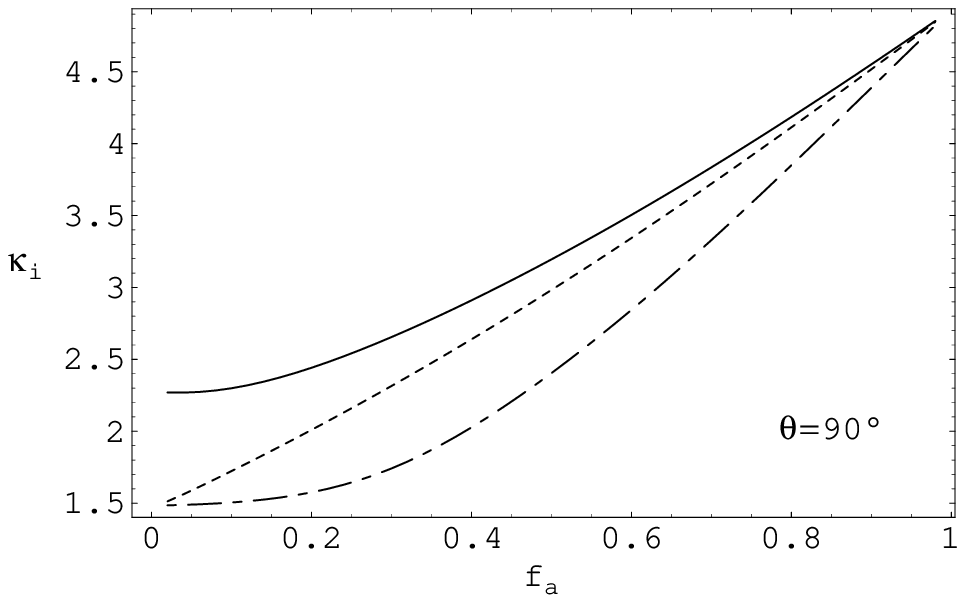,width=4.0in} \\
\epsfig{file=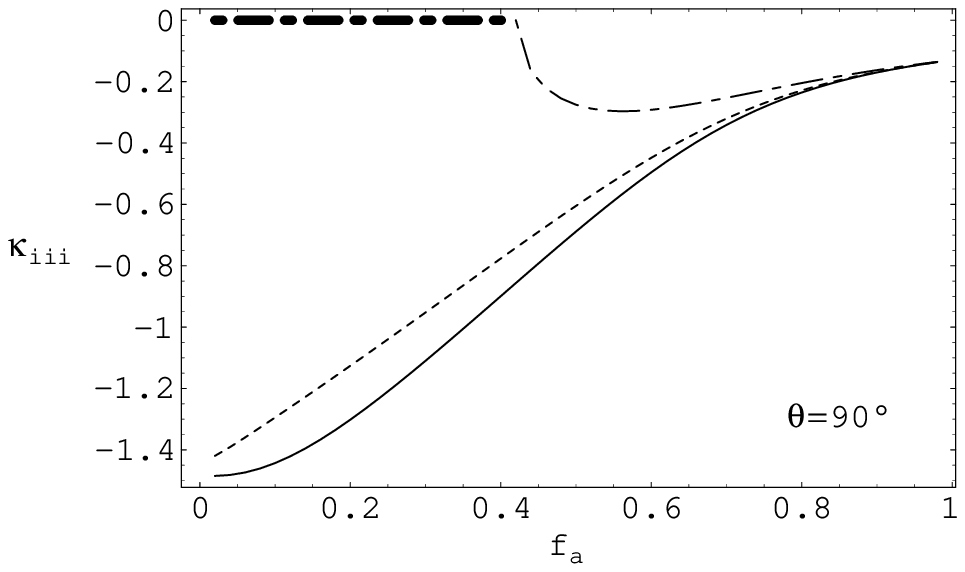,width=4.0in}
  \caption{\label{fig5}
  Same as figure \ref{fig3}, but for
relative wavenumbers
$\kappa_{i,iii}$ when $\theta=\pi/2$.}
\end{figure}

\newpage

\begin{figure}[!ht]
\centering \psfull \epsfig{file=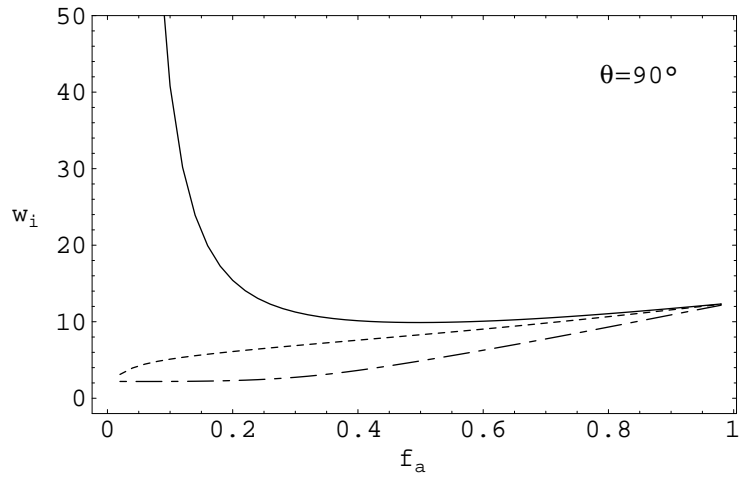,width=4.0in} \\
\epsfig{file=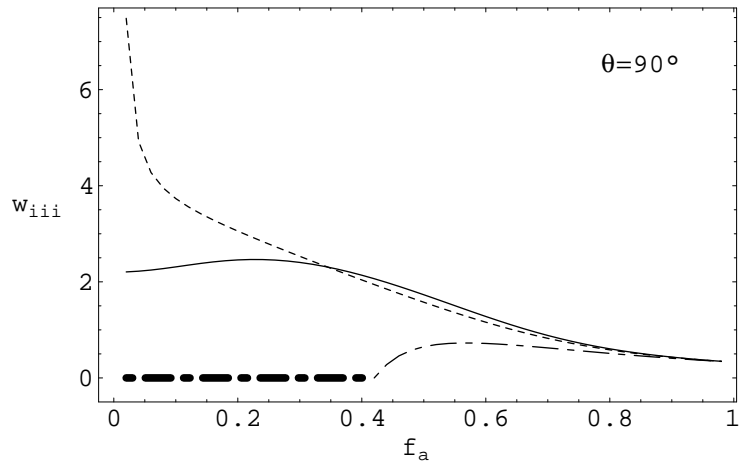,width=4.0in} \\
  \caption{\label{fig6}
  Same as figure \ref{fig4}, but for
$w_{i,iii}$ when $\theta=\pi/2$.
}

\end{figure}

\newpage

\begin{figure}[!ht]
\centering \psfull \epsfig{file=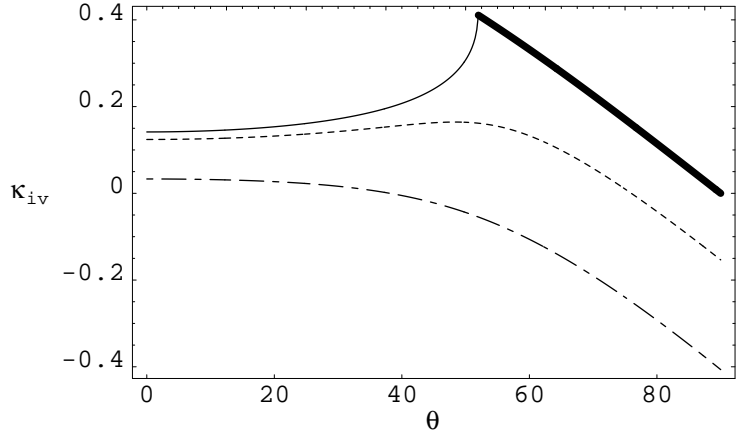,width=4.0in}
  \caption{\label{fig7}
Real parts of  relative wavenumbers $\kappa_{iv}$
as  functions of $\theta$ (in degrees) for the
dissipation parameter
$\delta
= 0, 0.1$ and $0.2$  when $f_a =0.35$.
The constitutive parameters of the
component mediums are:  $\epsilon^a = 3.2$, $ \xi^a = 2.4$, $ \mu^a = 2$,  $
\epsilon^b = 2.2 + i \, \delta$,
$\mu^b = 3.5 + i \, \delta$, $ \mu^b_z = 1 + i \, 0.5 \delta$,
and  $ \mu^b_g = 4 + i \, 2 \delta$.
  Key:  $\kappa_{iv}$ values corresponding to $\delta =
0, 0.1$ and $0.2$ are represented
by the solid lines, dashed lines,
and broken dashed  lines, respectively.
The heavy line on the graph for $\delta = 0$ 
indicates those relative
wavenumbers $\kappa_{iv}$ which  have
 nonzero imaginary parts.
}
\end{figure}

\vspace{10mm}

\begin{figure}[!ht]
\centering \psfull
\epsfig{file=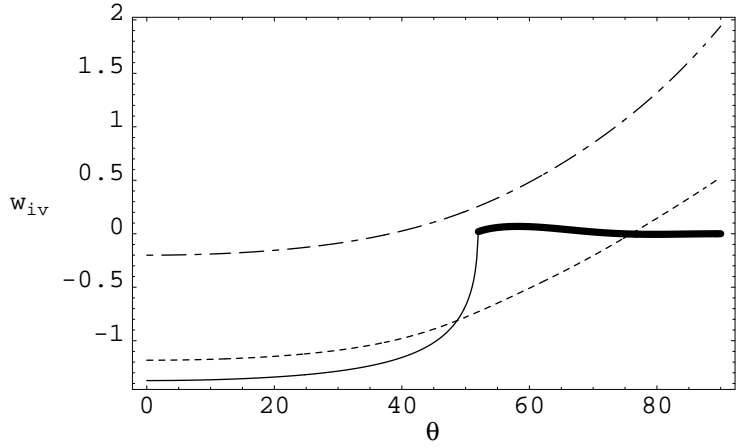,width=4.0in}
  \caption{\label{fig8}
Calculated values of $w_{iv}$ 
as  functions of $\theta$ (in degrees) for  the
dissipation parameter
$\delta
= 0, 0.1$ and $0.2$ when  $f_a =0.35$.
The constitutive parameters of the
component mediums are as in figure \ref{fig7}.
 Key:  $w_{iv}$ values corresponding to $\delta =
0, 0.1$ and $0.2$ are represented
by the solid lines, dashed lines,
and broken dashed  lines, respectively.
The heavy line on the graph for $\delta = 0$ 
indicates those
 $w_{iv}$
values which devolve from relative
wavenumbers $\kappa_{iv}$ with
 nonzero imaginary parts.
}
\end{figure}

\end{document}